# Numerical Simulation of Three-dimensional High-Lift Configurations Using Data-Driven Turbulence Model

Shaoguang Zhang,[1,*] Chenyu Wu,[1,†] Yufei Zhang[1,‡]
(*1Tsinghua University, Beijing, 100084, People's Republic of China*)

**Traditional Reynolds-averaged Navier-Stokes (RANS) equations often struggle to predict separated flows accurately. Recent studies have employed data-driven methods to enhance predictions by modifying baseline equations, such as field inversion and machine learning (FIML) with symbolic regression. However, data-driven turbulence models exhibit limited adaptability and are rarely applied to complex engineering problems. This study examines the application of data-driven turbulence models to complex three-dimensional high-lift configurations, extending their usability beyond previous applications. First, the generalizability of the SST-CND model, derived from conditioned field inversion and symbolic regression, is validated. Then, the spatially varying correction factor obtained through conditioned field inversion is transferred to the three-equation $k - \overline{v^2} - \omega$ model. The 30P30N three-element airfoil, the JAXA Standard Model (JSM), and the high-lift version of the NASA Common Research Model (CRM-HL) are numerically simulated. The results indicated that the SST-CND model significantly improves the prediction of stall characteristics, demonstrating satisfactory generalizability. The corrected $k - \overline{v^2} - \omega - $ CND model accurately predicts the stall characteristics of CRM-HL, with a relative error of less than 5% compared to experimental results. This confirms the strong transferability of the model correction derived from conditioned field inversion across different turbulence models.**

[*] Postdoctoral research assistant, School of Aerospace Engineering, email: zsg26@mail.tsinghua.edu.cn
[†] Ph. D student, School of Aerospace Engineering, email: wcy22@mails.tsinghua.edu.cn
[‡] Associate professor, School of Aerospace Engineering, senior member AIAA, email: zhangyufei@tsinghua.edu.cn (corresponding author)

**Nomenclature**

(Nomenclature entries should have the units identified)

| | | |
|---|---|---|
| $C_L$ | = | lift coefficient |
| $C_m$ | = | pitching moment coefficient |
| $C_p$ | = | pressure coefficient |
| $\rho$ | = | air density, kg/m$^3$ |
| $Re$ | = | Reynolds number |
| $S$ | = | shear rate, 1/s |
| $\mu$ | = | molecular viscosity, Pa.s |
| $\nu$ | = | kinematic molecular viscosity, $\mu/\rho$, $m^2/s$ |
| $\Omega$ | = | magnitude of vorticity, $1/s$ |
| $C_{pt}$ | = | stagnation pressure coefficient |

# I. Introduction

An accurate turbulence model is crucial for engineering applications. Among various turbulence modeling methods, Reynolds-averaged Navier-Stokes (RANS) equations are widely applied in engineering tasks due to their high computational efficiency compared to direct numerical simulation and large eddy simulation. Duraisamy et al. [1] argued that the RANS method will continue to serve computational fluid dynamics (CFD) as a fundamental approach to turbulence modeling for an extended period. The RANS method relies on turbulence models to describe turbulent motion. However, current turbulence models often fail in complex separated flows due to various simplifications and assumptions made during the model construction stage [2],[3]. These limitations restrict the application of the RANS method in engineering designs.

In recent years, data-driven methods have been widely utilized in turbulence modeling [3]. Among these, machine learning (ML) models have emerged as a significant advancement in traditional CFD, particularly in addressing the challenges of predicting separated flows using RANS methods [4–9]. Duraisamy et al. [10] demonstrated the potential of the field inversion and machine learning (FIML) method to quantify and mitigate the deficiencies of the RANS method using only sparse high-fidelity simulation data or wind tunnel measurement data. Extensive research has confirmed the successful application of FIML in modeling separated flows. Singh et al. [11] improved the Spalart-Allmaras (SA) model's capability in predicting wind turbine airfoil stall performance by employing a multilayer perceptron (MLP) model with FIML. Similarly, Yan et al. [12,13] enhanced the SA model's ability to predict flow around an iced airfoil and the three-dimensional separated flow around NASA's FAITH hill [14] using FIML with an MLP model. Although the FIML model significantly enhances accuracy for flows similar to those in its training datasets, concerns regarding its generalizability remain. Wu et al. [15] categorized the generalizability of the FIML model into four levels: Level 1 (L1) denotes that the model performs well on geometries similar to the training set; Level 2 (L2) indicates that the baseline model's accuracy remains unaffected for simple attached flows; Level 3 (L3) indicates that the model effectively predicts flows with separation characteristics similar to the training set but under different geometries and Reynolds numbers; Level 4 (L4) represents that the model performs robustly across various test cases involving different flow separation features, geometries, and Reynolds numbers. Previous studies have predominantly concentrated on L1 generalizability, with some achieving L3 generalizability. However, L2 generalizability is crucial for the baseline model. Wu et al. [15] developed the SST-CND model using a conditioned field inversion method, enhancing its L2 generalizability while preserving L1 and L3 generalizability. Nonetheless, attaining L4 generalizability remains a formidable challenge in machine learning-based RANS modeling and is included in NASA's five-year roadmap for CFD Vision 2030 [16]. The current literature indicates that applications of data-driven turbulence models are primarily confined to two-dimensional separated flows in simple geometries [6],[7],[17],[18], such as periodic hills. Some studies primarily investigate separation induced by adverse pressure gradients on airfoils [12],[19],[20]. A limited number of studies have attempted to validate data-driven models in relatively simple three-dimensional flows [13],[21][22]. The application of data-driven turbulence models to complex real-world engineering scenarios remains rare in the literature.

This study extends the boundary of the application of the data-driven turbulence model by applying it to complex three-dimensional high-lift devices. The aerodynamic performance of high-lift configurations is crucial for ensuring the safety and efficiency of commercial aircraft. The maximum lift coefficient plays a key role in optimizing high-lift designs and directly affects the payload capacity of the aircraft [23]. However, the flow structure of high-lift configurations is highly complex, involving phenomena such as transition, wake and boundary layer merging, and possible boundary layer separation in multi-element airfoil flows [24]. Flow separation in high-lift configurations often involves trailing-edge separation of the flap at small angles of attack with large deflections, as well as separation regions above the flap at high angles of attack. These complex flow phenomena make it challenging to accurately predict the lift increment caused by flap deflection and stall performance, posing significant challenges in high-lift configuration design [25].

Fast and accurate aerodynamic prediction methods are required for optimizing high-lift configurations. Currently, the Reynolds-Averaged Navier-Stokes (RANS) method remains widely used in the optimization and design of high-lift configurations. However, conclusions from the High-Lift Prediction Workshop organized by the American Institute of Aeronautics and Astronautics indicate that current RANS methods still fail to accurately predict forces and moments near the stall angle of attack [26–28]. RANS also fails to predict flap deflection effects at low angles of attack away from the stall. Both situations involve significant regions of separated flow, which represent a critical limitation of RANS [25].

This study applies the data-driven turbulence model produced by conditioned field inversion in [15] (SST-CND model) to high-lift configurations. First, the typical multi-element airfoil 30P30N, the JAXA Standard Model (JSM), and the high-lift version of the NASA Common Research Model (CRM-HL) are selected as test cases to validate the generalization capability of the SST-CND model in aerodynamic predictions for high-lift configurations. Then, the correction derived from conditioned field inversion is applied to the three-equation $k - \overline{v^2} - \omega$ transition model, creating the $k - \overline{v^2} - \omega - \text{CND}$ model. The accuracy of the $k - \overline{v^2} - \omega - \text{CND}$ model in predicting aerodynamic performance is validated using the CRM-HL case, and the mechanism of model correction is analyzed in detail. The results indicate that the SST-CND model demonstrates satisfactory generalizability in the 30P30N and JSM, while the correction derived from conditioned field inversion is transferable to the $k - \overline{v^2} - \omega$ turbulence model and exhibits high accuracy in the CRM-HL model. These results also demonstrate that properly trained data-driven turbulence models can be applied to complex real-world engineering flows.

## II. Numerical Method

### A. Numerical Solver

CFL3D version 6.7 [29] is employed in this study to predict the stall behavior of the high-lift configuration. Spatial discretization is performed using Roe's [30] flux-difference splitting technique, while van Leer's [31] Monotone Upstream-Centered Scheme for Conservation Laws (MUSCL) approach is applied for state-variable interpolation at cell interfaces. Time advancement is achieved through the implicit approximate-factorization method, with multigrid and mesh sequencing employed to accelerate convergence. This solver supports multiple-zone grids connected in one-to-one, patched, or overset manners.

**B. Turbulence Model**

1. SST-CND model

The turbulence model used in this study is derived from conditioned field inversion (FI-CND) [15] based on the SST 2003 model [32]. The transport equations of the SST model are as follows:

$$\frac{\partial(\rho k)}{\partial t} + u_j\frac{\partial(\rho u_j k)}{\partial x_j} = P - \beta^* \rho\omega k + \frac{\partial}{\partial x_j}\left[(\mu + \sigma_k\mu_t)\frac{\partial k}{\partial x_j}\right]$$

$$\frac{\partial(\rho\omega)}{\partial t} + \frac{\partial(\rho u_j\omega)}{\partial x_j} = \frac{\gamma}{\nu_t}P - \theta\rho\omega^2 + \frac{\partial}{\partial x_j}[(\mu + \sigma_\omega\mu_T)]\frac{\partial\omega}{\partial x_j} + 2(1-F_1)\frac{\rho\sigma_{\omega 2}}{\omega}\frac{\partial k}{\partial x_j}\frac{\partial\omega}{\partial x_j}$$

(1)

where $k$ is the turbulent kinetic energy, $\omega$ denotes the specific dissipation rate, and $\mu_t$ is the eddy viscosity. For a detailed explanation of the SST model, refer to reference [32].

A spatially varying correction factor $\beta(\mathbf{X})$ ($\mathbf{X}$ is the spatial coordinate) in classical field inversion is directly multiplied by the destruction term of the $\omega$ equation. This allows the correction factor to vary at any location within the flow field, including the attached boundary layer. Although the classical field inversion (FI-CLS) can accurately predict the reattachment point in the NASA hump case [33], it tends to overestimate the friction coefficient at the top of the hump. In addition, the model derived by FI-CLS is likely to negatively affect the accuracy of zero pressure-gradient flat plates, which is already nicely treated by the baseline SST 2003 model. Therefore, Wu et al. [15] proposed the FI-CND method, where the multiplier form of the destruction term in the $\omega$ equation is expressed as shown in Eq. (2):

$$\frac{\partial(\rho\omega)}{\partial t} + \frac{\partial(\rho u_j\omega)}{\partial x_j} = \frac{\gamma}{\mu_t}P - \beta\theta\rho\omega^2 + \frac{\partial}{\partial x_j}[(\mu + \sigma_\omega\mu_T)]\frac{\partial\omega}{\partial x_j}$$
$$+ 2(1-F_1)\frac{\rho\sigma_{\omega 2}}{\omega}\frac{\partial k}{\partial x_j}\frac{\partial\omega}{\partial x_j}$$

(2)

where $\beta$ is defined as $\beta = [(B(\mathbf{X}) - 1) * f_d + 1]$. When $B(\mathbf{X}) = 1$, Eq. (2) can reduce to the baseline model's $\omega$ equation. The shielding function proposed by Spalart et al. [34], is expressed as follows:

$$f_d = 1 - \tanh[(8r_d)^3], r_d = \frac{\mu + \mu_T}{\rho \kappa^2 d^2 \sqrt{u_{i,j} u_{i,j}}} \tag{3}$$

where $d$ is the wall distance and $u_{i,j} = \partial u_i / \partial x_j$. $f_d$ is constructed such that it is 0 inside the boundary layer and 1 outside. This means that within the boundary layer, the model aligns with the baseline SST model. Outside the shielding region, the correction factor modifies the destruction term to $B(\mathbf{X})\theta\rho\omega^2$. For a detailed description, readers can refer to the literature [15].

A simple introduction to the field inversion process is included. In the field inversion, the value of $\beta$ on each cell is obtained by solving an optimization problem defined in Eq. (4):

$$\min_{B} J = \lambda_{QoI} \sum_{i=1}^{K} [d_i - h_i(\mathbf{B})]^2 + \lambda_{L_2} \sum_{j=1}^{N} (B_j - 1)^2 \tag{4}$$

where $\mathbf{B}$ is a vector whose $j^{th}$ value is $B_j$. $B_j$ is the value of the correction term $B(\mathbf{X})$ in the $j^{th}$ cell. $d_i$ is the $i^{th}$ high-fidelity data and $h_i$ is the value predicted by RANS. $\lambda_{QoI}$ and $\lambda_{L_2}$ are two positive constants given by the user. The first term in Eq. (4) indicates that the error between the RANS prediction and the high-fidelity data is minimized by adjusting $\mathbf{B}$. The second term means that $\beta$ is limited from deviating its default value too far. The optimization problem in Eq. (4) is solved using a gradient-based algorithm. The gradient is computed by a discrete adjoint method implemented by DAFoam [17],[35-37]. All the CFD calculations in the field inversion process are done using OpenFOAM [38].

The training process of the SST-CND model is illustrated in Fig. 1. First, FI-CND is performed on both the NASA hump case ($Re \approx 1 \times 10^6$) [33] and the curved backward-facing step ($Re \approx 1 \times 10^4$) [39] case. The $x$-directional velocity given by LES [39],[40] is used as the high-fidelity data. The location of the high-fidelity data points is marked by blue triangles in the first row of Fig. 1. Conditioned field inversion is conducted to produce the spatial distribution of $\beta$ shown in the second row of Fig. 1. The datasets of $\beta$ generated by field inversion of the two cases are then combined to capture the characteristics of both medium Reynolds number and low Reynolds number. At last, symbolic regression is conducted using PySR [41] on the combined dataset. The final expression obtained is shown in Eq. (5).

$$B(\mathbf{X}) - 1 = \min(0.00435\lambda_2^2, 3.806) \tag{5}$$

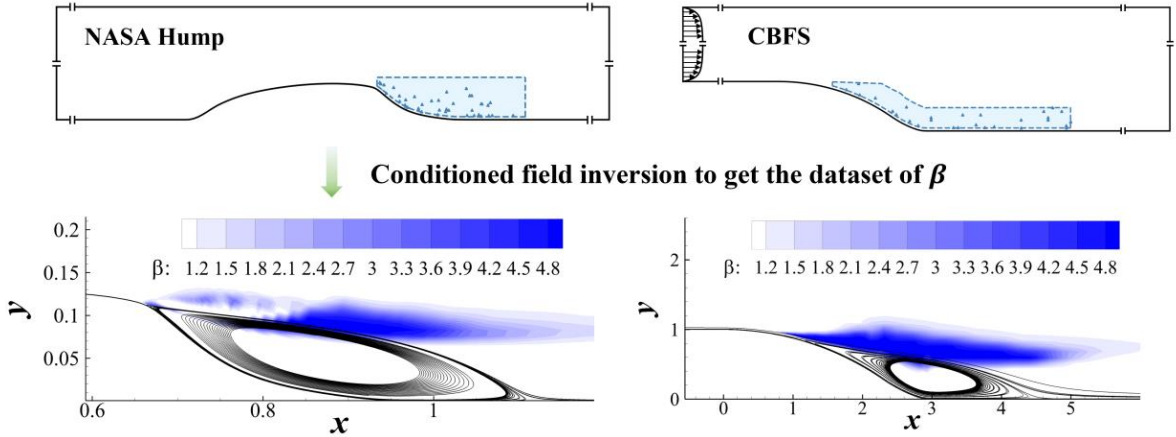

Fig. 1 The training process of the model [15]

where $\lambda_2 = tr(\hat{\Omega}^2)$ is the 2nd scaler invariances of nondimensional strain rate $\hat{S} = S/(\beta^*\omega)$ and nondimensional rotation rate $\hat{\Omega} = \Omega/(\beta^*\omega)$ derived by Pope [42] for general tensor representation for the Reynolds stress. $\beta^*$ is a model constant of the SST model and equals 0.09.

The SST-CND model demonstrates strong generalizability in typical separated flow cases, including the NLR7301 multi-element airfoil and the Ahmed body [15], as illustrated in Fig. 2. This study applies the SST-CND model to complex high-lift configurations to further assess the generalizability of the current data-driven turbulence model.

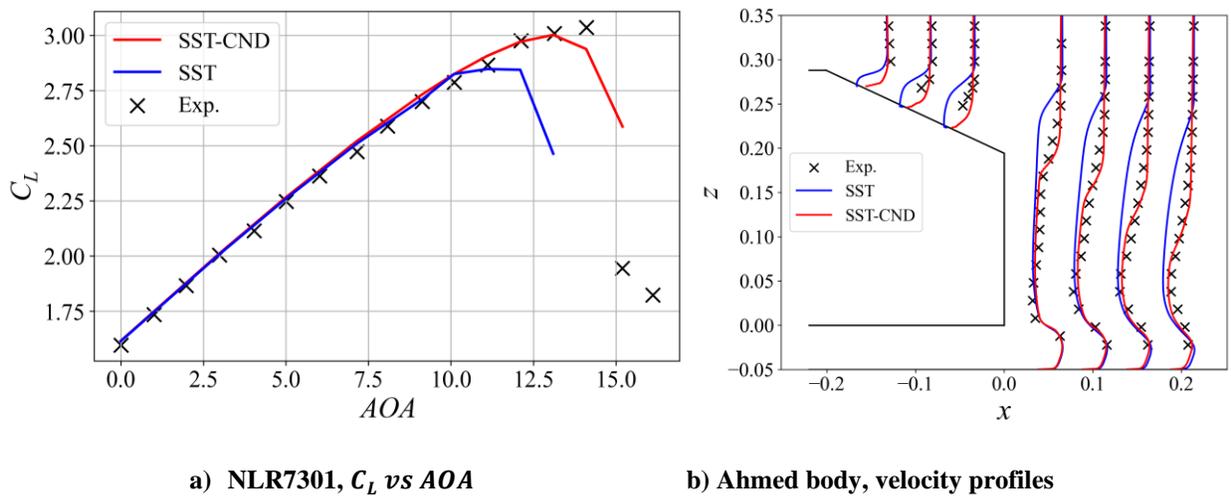

a) NLR7301, $C_L$ vs $AOA$  b) Ahmed body, velocity profiles

Fig. 2 The test cases of the SST-CND model in Ref [15]

2. $k - \overline{v^2} - \omega -$ CND model

The spatially varying correction factor $\beta$ obtained through FI-CND has a concise form, facilitating its transferability. Therefore, it is applied to the $k - \overline{v^2} - \omega$ model, which has been validated in previous research for its effectiveness in predicting stall behavior in high-lift configurations [43]. The original $k - \overline{v^2} - \omega$ model was initially proposed by Lopez and Walters in 2016 [44] and consists of three transport equations: total fluctuation energy ($k$), fully turbulent fluctuation energy ($\overline{v^2}$), and specific dissipation rate ($\omega$), as expressed in Eqs. (7), (8), and (9). For detailed information on each term of this model, readers are referred to reference [45]. The modification is implemented by applying the coefficient $\beta$ in Eq. (8) to the destruction term of the $\omega$ transport equation. The form of $\beta$ in Eq. (8) is the same as that in Eq. (2). The modified model is named the $k - \overline{v^2} - \omega -$ CND model.

$$\frac{\partial k}{\partial t} + u_j \frac{\partial k}{\partial x_j} = \frac{1}{\rho} P_k - min(\omega k, \omega \overline{v^2}) - \frac{1}{\rho} D_k + \frac{1}{\rho} \frac{\partial}{\partial x_j} \left[ \left( \mu + \frac{\rho \alpha_T}{\sigma_k} \right) \frac{\partial k}{\partial x_j} \right] \quad (6)$$

$$\frac{\partial \overline{v^2}}{\partial t} + u_j \frac{\partial \overline{v^2}}{\partial x_j} = \frac{1}{\rho} P_{\overline{v^2}} + R_{BP} + R_{NAT} - \omega \overline{v^2} - \frac{1}{\rho} D_{\overline{v^2}} + \frac{1}{\rho} \frac{\partial}{\partial x_j} \left[ \left( \mu + \frac{\rho \alpha_T}{\sigma_k} \right) \frac{\partial \overline{v^2}}{\partial x_j} \right] \quad (7)$$

$$\frac{\partial \omega}{\partial t} + u_j \frac{\partial \omega}{\partial x_j} = \frac{1}{\rho} P_\omega + \left( \frac{C_{\omega R}}{f_W} - 1 \right) \frac{\omega}{\overline{v^2}} (R_{BP} + R_{NAT}) - \beta C_{\omega 2} \omega^2 f_W^2$$
$$+ 2\beta^*(1 - F_1^*) \sigma_{\omega 2} \frac{1}{\omega} \frac{\partial k}{\partial x_j} \frac{\partial \omega}{\partial x_j} + \frac{1}{\rho} \frac{\partial}{\partial x_j} \left[ \left( \mu + \frac{\rho \alpha_T}{\sigma_\omega} \right) \frac{\partial \omega}{\partial x_j} \right] \quad (8)$$

### III. Test Cases

**A. 30P30N Multielement Airfoil**

The 30P30N multi-element airfoil is selected as the first test case to evaluate the accuracy of the SST-CND model in predicting stall performance. This multi-element airfoil details experimental data provided by NASA Langley Research Center's Low Turbulence Pressure Tunnel (LTPT) [46], including aerodynamic coefficients and pressure distribution. The freestream conditions are set to $M_\infty = 0.2$ and $Re_C = 9 \times 10^6$. The mesh utilized in this study is identical to that in the literature [47], with refinement applied to the upper surfaces of the main wing and flap, as shown in Fig. 3. The computational grid consists of 246,000 cells, with the first grid layer set at the height of $5.0 \times 10^{-6}$ to ensure that $\Delta y^+$ remains below 1.0.

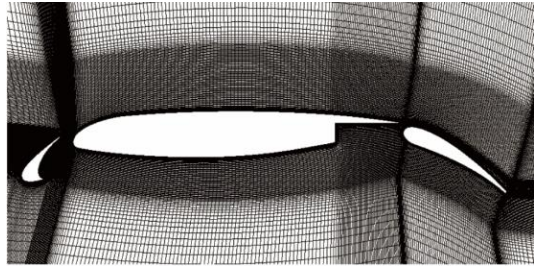

**Fig. 3 Computational grid of the 30P30N multi-element airfoil**

Aerodynamic coefficients predicted by the SST and SST-CND models are presented in Fig. 4. The SST model underestimates the maximum lift coefficient by 3.45%, primarily due to its underestimation of the loads on the main wing and flap, as shown in Fig. 4(b) and Fig. 4(c). In contrast, the SST-CND model slightly overestimates the maximum lift coefficient, yielding a relative error of 1.61%. The experimental stall angle of attack is 21°, while both the SST and SST-CND models predict a 2° delay, as indicated in Table 1. Fig. 4(d) presents the pressure coefficient ($C_p$) distribution at $AOA = 21°$. The SST model underpredicts the suction peak of the main wing, whereas the SST-CND model exhibits a stronger correlation with the experimental data. Fig. 5 displays the nondimensional streamwise velocity $U/U_{inf}$ contours predicted by the SST and SST-CND models at $AOA = 21°$. The two low-speed regions above the flap correspond to the slat wake and the main wing wake. These regions generate shear layers both above and below due to significant velocity gradients. The largest differences in velocity contours predicted by the turbulence models occur within the wake region. Flow reversal above the flap is predicted by the SST model, leading to an overprediction of the wake width and an underestimation of velocity within the wake area, as shown in Fig. 6. The locations for surface-normal profile measurements in Fig. 6 are detailed in reference [48]. The SST-CND model predicts a smaller, more localized main wing wake near the flap surface compared to the SST model but underestimates the velocity of the slat wake.

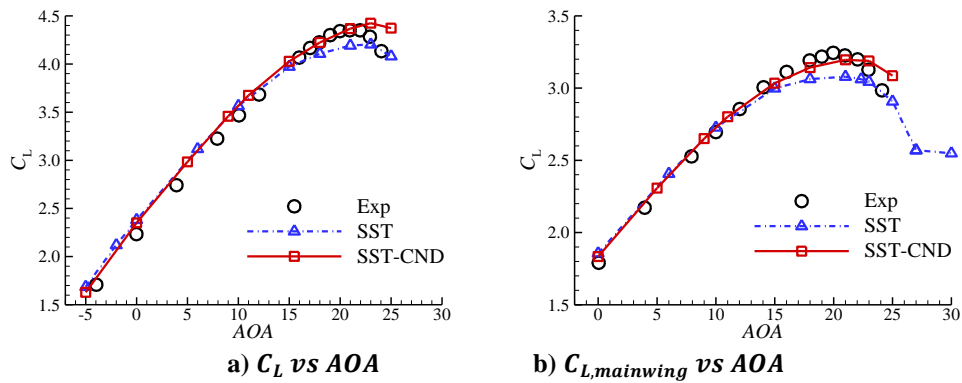

a) $C_L$ vs AOA  b) $C_{L,mainwing}$ vs AOA

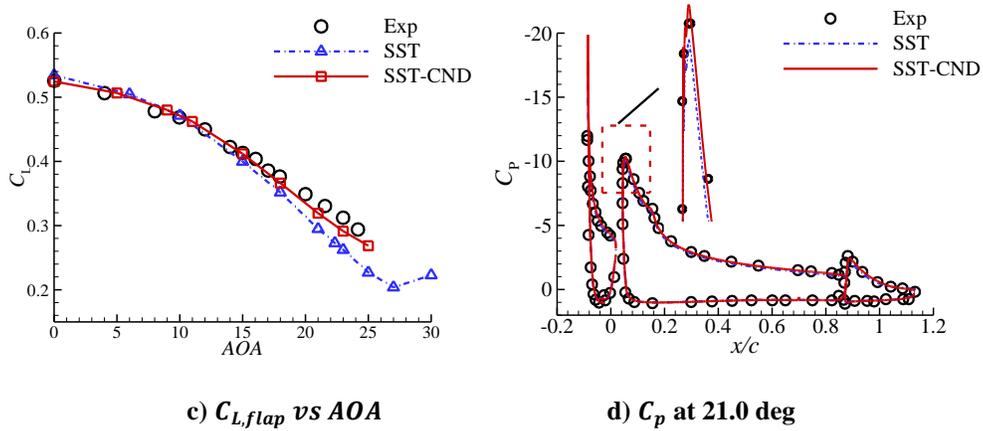

c) $C_{L,flap}$ vs AOA

d) $C_p$ at 21.0 deg

**Fig. 4 The aerodynamic performance given by the SST and SST-CND models**

**Table 1. The results and relative errors of different models**

|  | Experiment | SST | SST-CND |
| --- | --- | --- | --- |
| $C_{L,max}$/relative error | 4.35/-- | 4.20/3.45% | 4.42/1.61% |
| Stall AOA/deviation | 21°/-- | 23°/2° | 23°/2° |

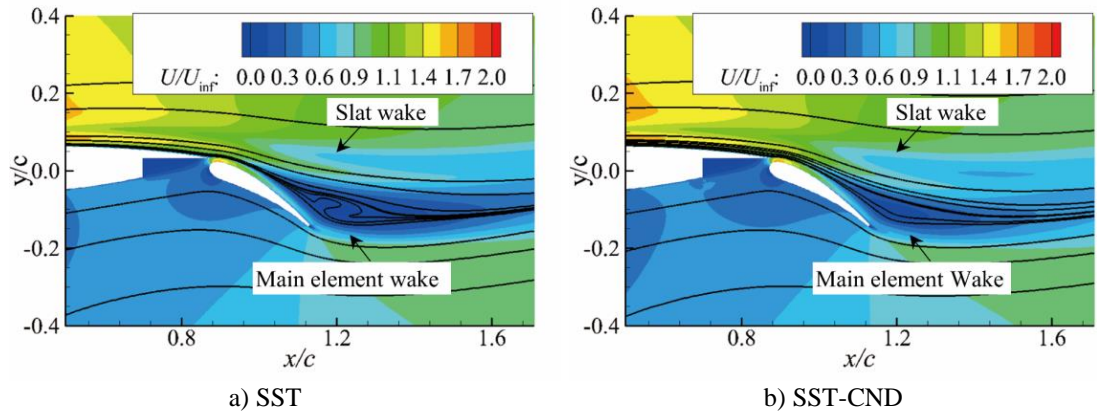

a) SST

b) SST-CND

**Fig. 5 Nondimensional streamwise velocity $U/U_{inf}$ contours predicted by different turbulence models, AOA=21.0 deg**

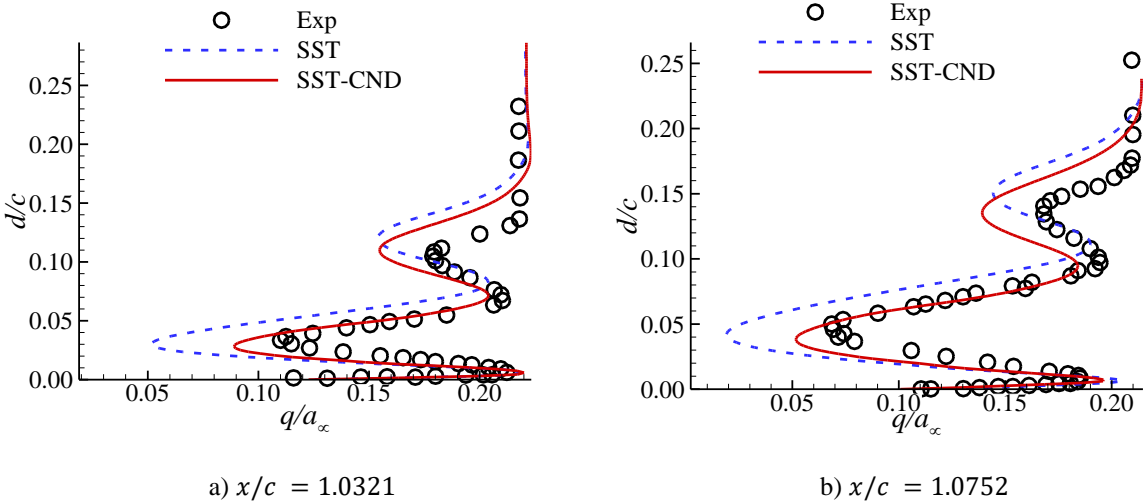

a) $x/c = 1.0321$  b) $x/c = 1.0752$

**Fig. 6 Velocity profile predicted by the SST and SST-CND models for $AOA = 21°$**

$\beta$ increases significantly near the mixing layers formed between the mainstream and wake, as well as between the wake and the gap jet, as depicted in Fig. 7. The increase in $\beta$ improves the destruction of $\omega$, reducing the dissipation of $k$, which rises turbulent viscosity ($\nu_T$). The increased $\nu_T$ enhances momentum diffusion from the mainstream to the shear layer compared to that modeled by the baseline SST model. Therefore, the SST-CND model predicts a smaller main wing wake than the baseline SST model.

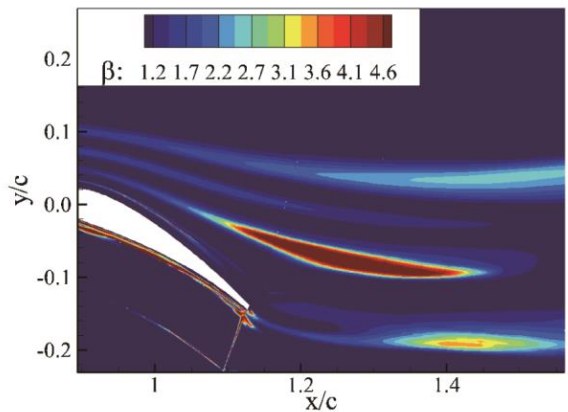

**Fig. 7 $\beta$ distributions obtained by SST-CND model**

### B. JAXA Standard Model

The JAXA Standard Model (JSM) was selected as the second test case to validate the accuracy of the SST-CND model in aerodynamic predictions for three-dimensional high-lift configurations. The JSM model was developed by JAXA and adopted by NASA as the standard model for the 3rd High-Lift Prediction Workshop (HiLiftPW-3) [28]. HiLiftPW-3 provides configurations both with (Case 2c) and without (Case 2a) nacelles/pylons. This study selects Case 2c, which includes nacelles/pylons. The slat covers 90% of the leading edge, and the configuration features a flap deflection angle of 30° and a slat deflection angle of 30°. The Reynolds number, based on the mean aerodynamic chord, is 1.93 million, with a freestream Mach number of 0.172. The tunnel turbulence intensity was estimated at 0.16%, and no transition trip was applied to the model.

The coarse, medium, and fine grids were based on the mesh from reference [43]. An extra fine grid was generated by increasing the number of grid points by 1.3 times in all three directions from the fine grid. The total grid counts for the four sets are 32, 56, 96, and 170 million, respectively. The wall grid counts are 0.41, 0.62, 1.32, and 2.52 million, respectively. The wall grid for the medium mesh is shown in Fig. 8, with the $x$ and $y$ directions representing the streamwise and spanwise directions.

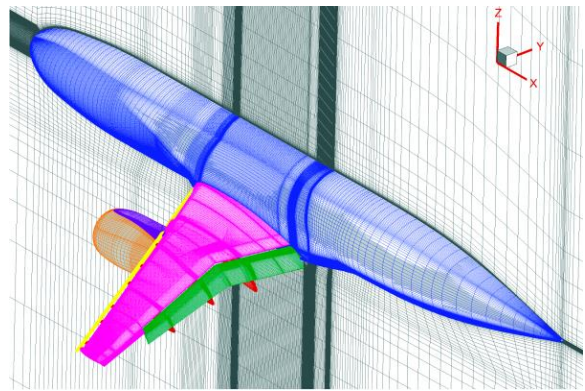

**Fig. 8 Wall grid of the JSM**

Fig. 9 presents the aerodynamic forces and moments of the JSM configuration predicted by the SST-CND model using the four sets of grids. The points marked with symbols on the curves in Fig. 9 represent all the computed conditions. Multigrid acceleration is applied to enhance the convergence of computations for each condition, typically involving 3000 computation steps on the first grid level, 2000 on the second, and at least 5000 on the third until convergence is reached. The scaled iterative convergence of $C_L$ for the SST-CND model with different grids at 10.48° and near-stall (17°) angles of attack achieves steady-state convergence, as shown in Fig. 10. The coarse grid underpredicts both the maximum lift coefficient and the stall angle of attack, as depicted in Fig. 9(a). The medium grid predicts a slightly higher maximum lift coefficient than the coarse grid, with no change in the stall angle. The fine grid not only predicts a higher maximum lift coefficient but also delays the stall angle from 17° to 18.59° compared to the medium grid. The results from the extra fine grid closely match those from the fine grid, confirming grid convergence. Therefore, all subsequent calculations are performed on the fine grid (with 96 million points) to ensure both accuracy and computational efficiency.

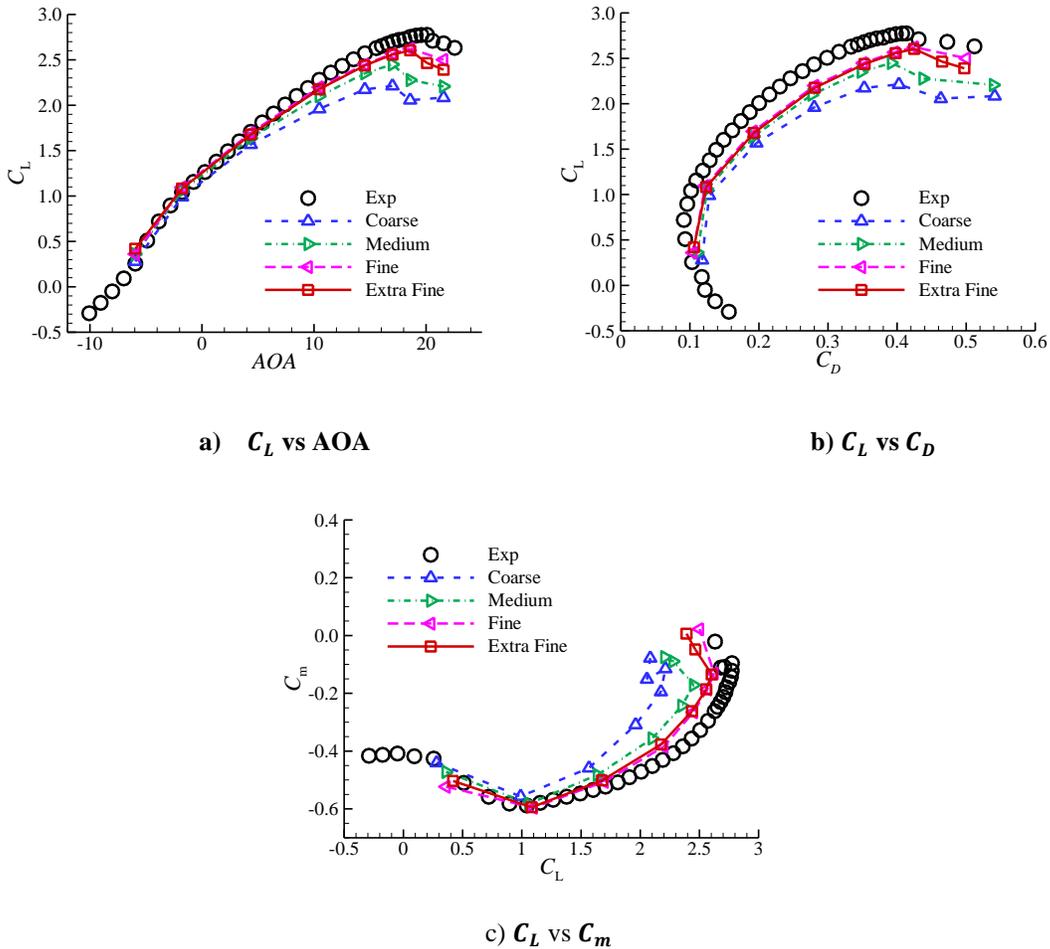

a) $C_L$ vs AOA

b) $C_L$ vs $C_D$

c) $C_L$ vs $C_m$

Fig. 9 Grid refinement study for the JSM configuration using the SST-CND model

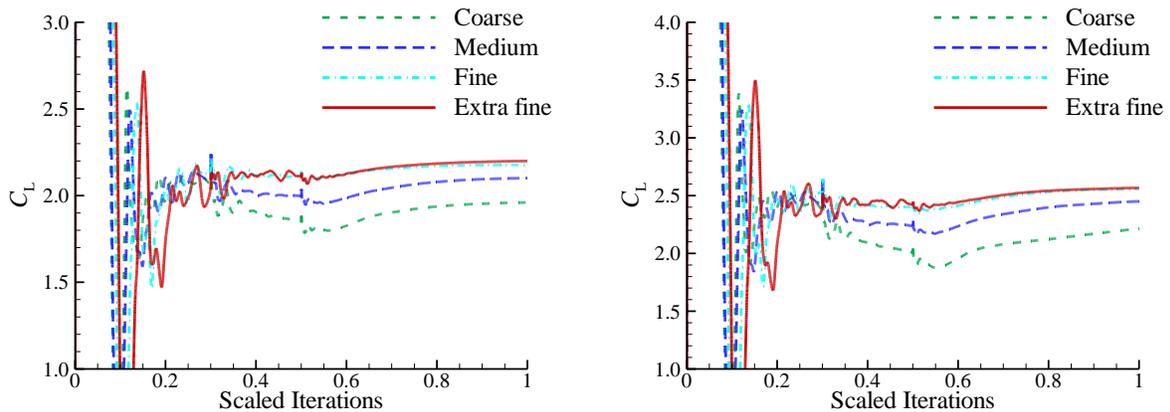

a) $AOA = 10.48°$         b) $AOA = 17.0°$

**Fig. 10 Scaled iterative convergence of $C_L$ for the SST-CND model with different grids**

Fig. 11 illustrates the aerodynamic performance of the JSM predicted by the SST and SST-CND models using fine grids. The lift curves reveal a significant difference in stall performance between the two models. The SST model underestimates the maximum lift coefficient by 17.69% and predicts the stall angle 5.55° earlier than the experimental value. In contrast, the SST-CND model reduces the relative error of the maximum lift coefficient to 6.13% and delays the stall angle to 18.59°, which remains 1.5° earlier than the experimental value, as listed in Table 2. In addition, the SST-CND model yields results that are more consistent with the experimental data for the polar and moment curves at the near-stall angle of attack, as depicted in Fig. 11(b) and Fig. 11(c).

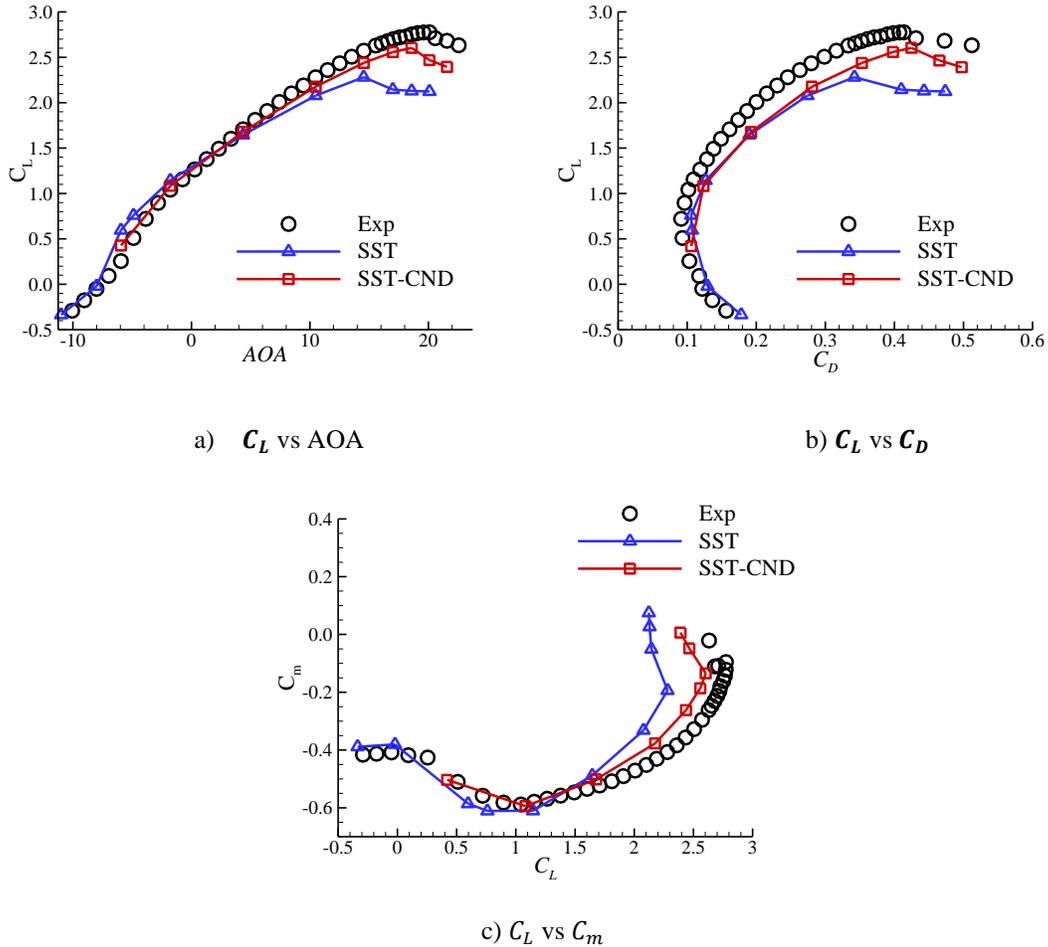

a) $C_L$ vs AOA

b) $C_L$ vs $C_D$

c) $C_L$ vs $C_m$

**Fig. 11 The aerodynamic performance of the JSM predicted by the SST and SST-CND models**

The experimental oil flow images at angles of attack of 18.58° and 21.57° are shown in Fig. 12. Significant flow separation is observed near the wingtip, with a larger separation at the wing root occurring at higher angles of attack. The tip separation appears to be induced by the disturbed flow from the main element directly behind the outermost slat bracket. The wake of each slat bracket is visible on the main wing. The SST model incorrectly predicts the influence of the slat wake, resulting in more pronounced tip separation and outer wing separation. Therefore, the pressure distribution on the trailing edge of the main wing predicted by the SST model exhibits a distinct plateau region, as shown in section E-E of Fig. 13. In contrast, the SST-CND model more accurately predicts stall characteristics and provides a more precise suction peak for both the slat and main wing, as illustrated in Fig. 13. The separation observed on the wing outboard aligns with the oil flow image and is effectively captured by the model.

**Table 2. The results and relative errors of the JSM predicted by the SST and SST-CND models**

|  | Experiment | SST | SST-CND |
| --- | --- | --- | --- |
| $C_{L,max}$/relative error | 2.77/-- | 2.28/17.69% | 2.60/6.13% |
| Stall AOA/deviation | 20.09°/-- | 14.54°/5.55° | 18.58°/1.50° |

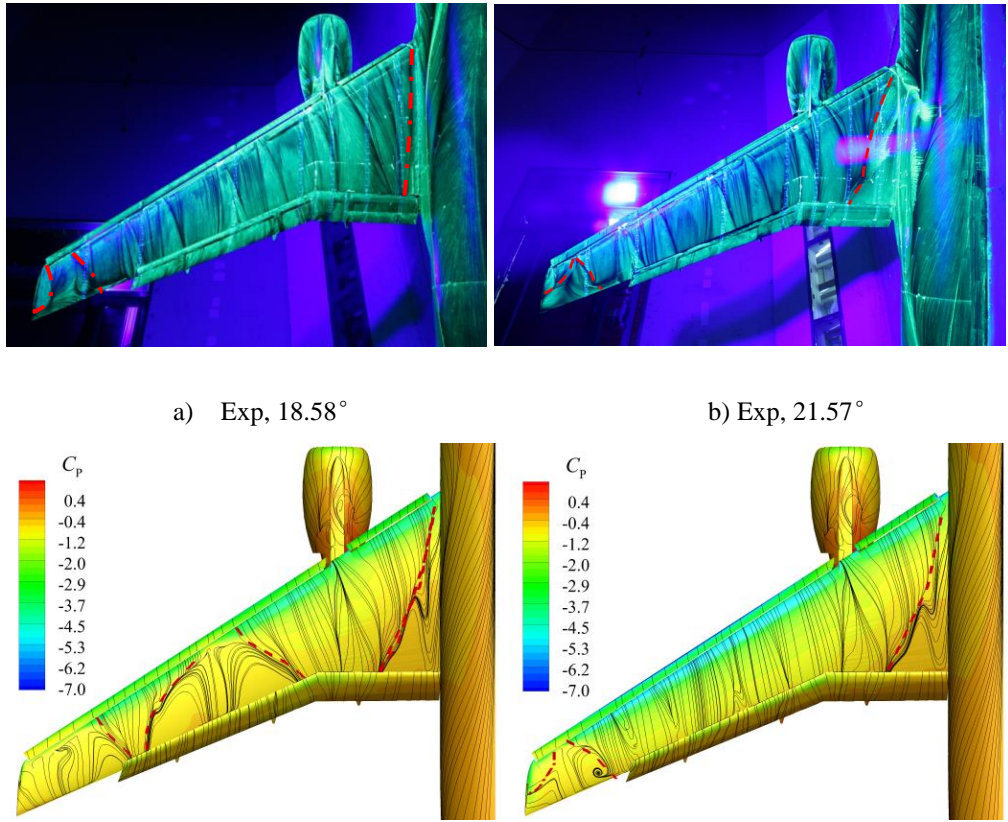

a) Exp, 18.58°  b) Exp, 21.57°

c) SST, 18.58°          d) SST-CND, 18.58°

**Fig. 12 The aerodynamic performance of the JSM predicted by the SST and SST-CND models**

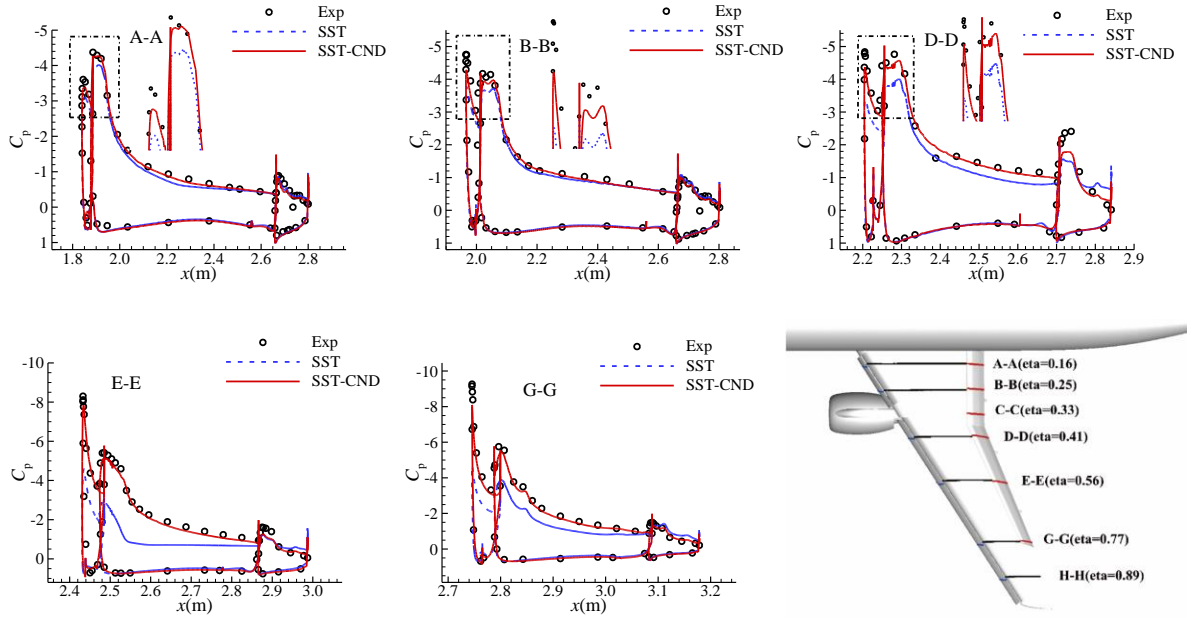

**Fig. 13 Pressure distributions of the JSM configuration at different span locations obtained by different models, $AOA = 18.58°$**

Fig. 14 indicates that for the SST-CND model predicting the results of JSM at $AOA = 18.58°$, $\beta$ increases in three key mixing regions, as depicted in Fig. 15. The first mixing region consists of the free shear layer formed by the interaction between the main wing wake and the mainstream flow. The second mixing region is the shear layer generated between the jet from the gap between the main wing trailing edge, the flap leading edge, and the main wing wake. The third mixing region constitutes a free shear layer formed at the interface between the low-velocity zone created by the merging of the main wing and flap wakes and the mainstream flow. In these mixed regions, an increase in $\beta$ enhances the destruction of $\omega$, reducing the dissipation of $k$, which in turn increases turbulent viscosity ($\nu_T$). The increased $\nu_T$, compared to that modeled by the baseline SST model, enhances momentum diffusion from the mainstream to the shear layer. The SST-CND model predicts smaller main wing wake regions, both inboard and outboard, as shown in Fig. 15(c) and Fig. 15(d). The reduction in wake increases the wing's circulation, improving the consistency between the predicted and experimental lift values.

16/35

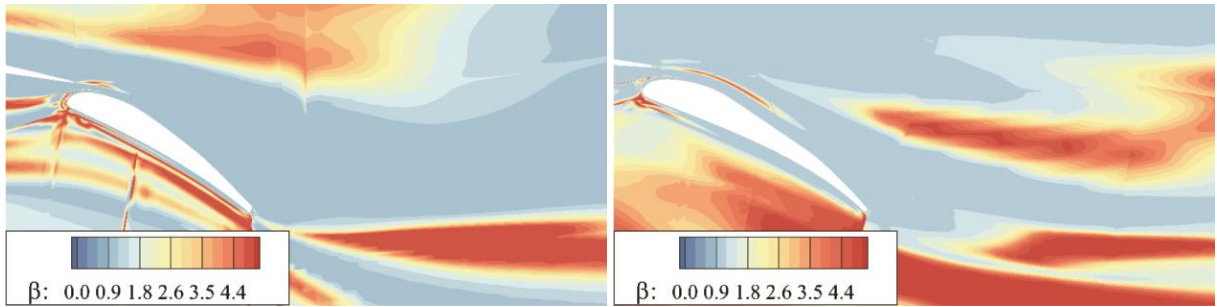

a) 17% span location          b) 70% span location

**Fig. 14 $\beta$ distributions obtained by the SST-CND model, AOA $= 18.58°$**

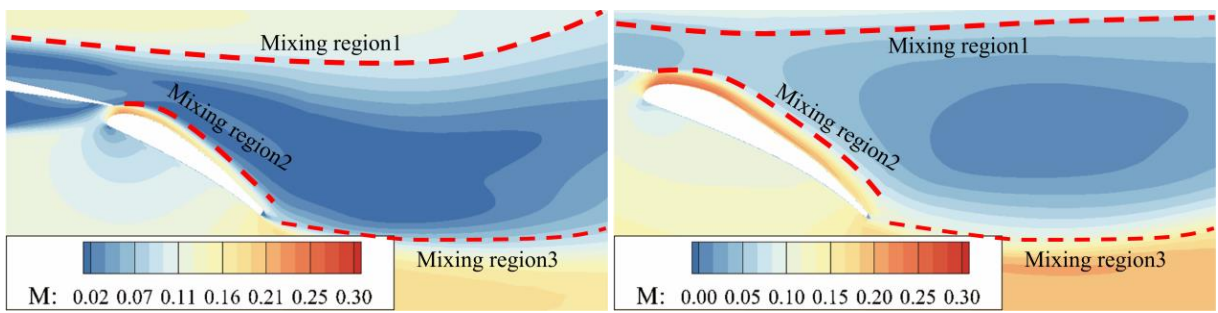

a) SST, 17% span location          b) SST, 70% span location

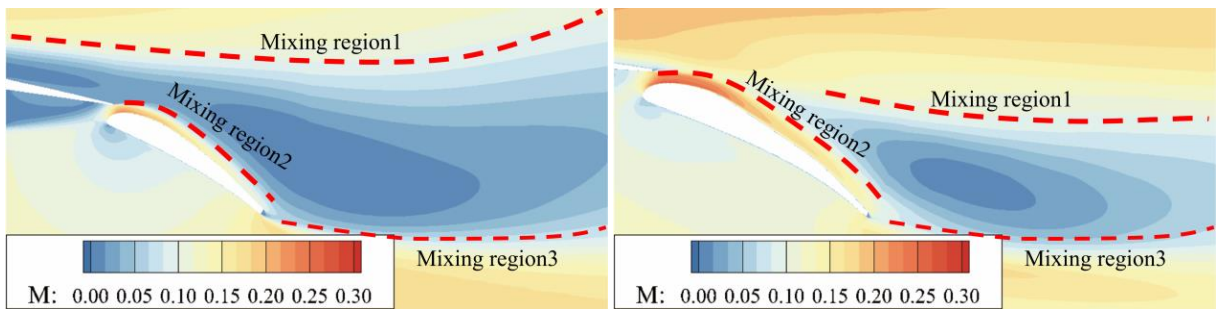

c) SST-CND, 17% span location          d) SST-CND, 70% span location

**Fig. 15 Mach number distribution contours obtained using different turbulence models, AOA $= 18.58°$**

### C. The High-lift Version of the Common Research Model

The CRM-HL model from the Fourth High-Lift Prediction Workshop is selected as the third test case [25]. Compared to the JSM model, the CRM-HL model features a higher Reynolds number, more complex geometry, and greater deviation from the training dataset of the data-driven turbulence model. This selection is intended to assess the generalization capability of the model. This section consists of two primary components. First, it evaluates the generalization capability of the SST-CND model, derived from FI-CND, in predicting the aerodynamic performance of CRM-HL. Then, it assesses the accuracy of the FI-CND correction applied to the $k - \overline{v^2} - \omega$ three-equation model for CRM-HL aerodynamic predictions.

*1. Geometry and Computational Grid*

The CRM-HL configuration closely represents a modern transport aircraft, incorporating components such as the fuselage, main wing, leading-edge slats, nacelle, nacelle pylon, trailing-edge flaps, and flap track fairings, as shown in (Fig. 16). Wind tunnel tests were conducted at the QinetiQ 5 m wind tunnel in Farnborough, where high-quality test data were provided. The workshop supplied three configurations with varying flap deflection angles. This study selects the nominal configuration [49] as the test case, with inboard and outboard flap deflection angles of 40° and 37°, respectively. The Reynolds number, based on the mean aerodynamic chord, is 5.49 million, with a freestream Mach number of 0.2. The mean longitudinal turbulence intensity of the incoming flow is 0.08%.

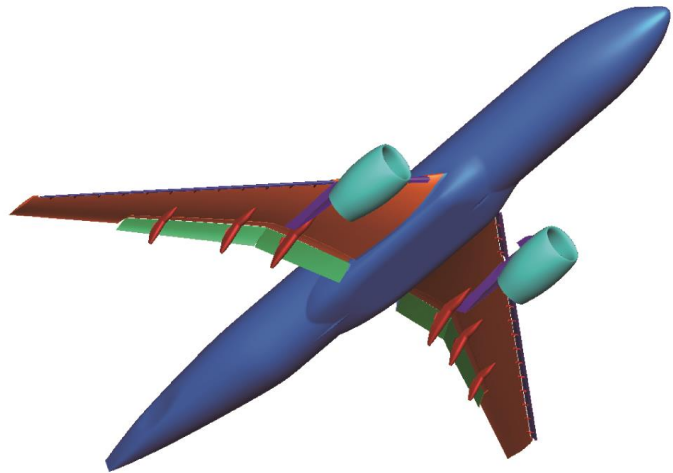

**Fig. 16 CRM-HL configuration**

The computation is based on structured grids from the previous study [50]. Four grid resolutions are utilized: coarse (78 million cells), medium (149 million cells), fine (240 million cells), and extra fine (493 million cells). The medium grid is generated by increasing the node count along each of the three edges by a factor of 1.3 compared to the coarse grid. The fine grid is generated using the same approach as the medium grid, while the extra fine grid is primarily refined in the flow and spanwise directions. Excessive refinement in the thickness direction can result in computational instability. Fig. 17 presents the wall grid of the medium resolution, where the $x$ and $y$ axes correspond to the streamwise and spanwise directions. The first grid layer of the four grid sets has $\Delta y^+$ values of approximately 1.0, 0.6, 0.4, and 0.2, respectively.

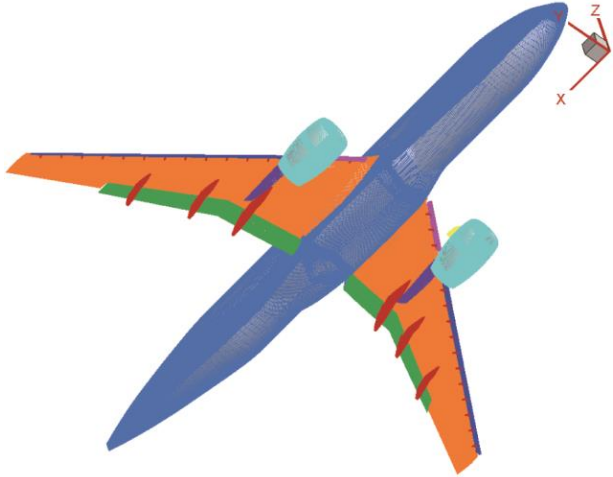

**Fig. 17 Wall grid of the CRM-HL medium grid**

*2. Results of the SST-CND Model*

    A grid convergence study is conducted on the lift, drag, and pitching moment coefficients using the SST-CND model, as illustrated in Fig. 18. The coarse and medium grids underestimate the maximum lift coefficient, whereas the fine grid provides a significant improvement. The fine and extra-fine grids yield nearly identical results at high angles of attack, demonstrating strong grid convergence. However, oscillatory convergence becomes more pronounced with the extra-fine grid, as shown in Fig. 18 (d). The coarse and medium grids tend to overestimate the drag coefficient. The fine and extra-fine grids produce results closer to experimental values but still overestimate drag near stall. The pitching moment coefficient results from the coarse, medium, and fine grids gradually approaching the experimental values as the grid is refined. However, the extra-fine grid results slightly deviate from the experimental values compared to the fine grid. Accordingly, the SST-CND model exhibits good grid convergence in predicting the aerodynamic performance of the CRM-HL configuration. The fine grid is employed for subsequent simulations to accurately assess differences in stall behavior across models.

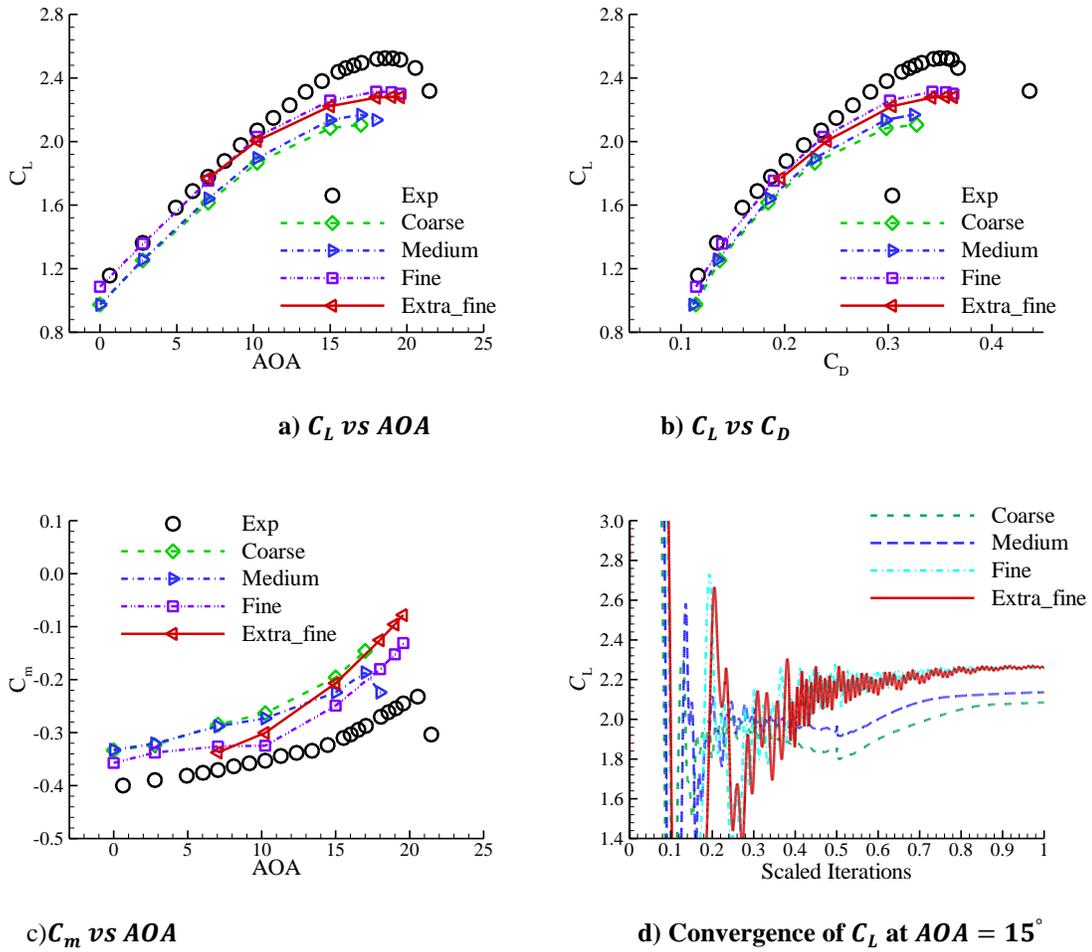

**Fig. 18 Mesh convergence of CRM-HL lift coefficient obtained by the SST-CND model**

The aerodynamic coefficients predicted by the SST and SST-CND models using fine grids are shown in Fig. 19. The SST model underestimates the maximum lift coefficient and introduces a nonlinear increase in the linear segment of the lift curve due to the underprediction of the lift coefficient at AOA = 7.05°. The wind tunnel test provides a corrected lift coefficient of 1.78 at AOA = 7.05°, whereas the SST and SST-CND models predict values of 1.69 and 1.75, respectively. The SST-CND model outperforms the SST model by providing a more accurate lift coefficient in the linear segment and a maximum lift coefficient closer to experimental data. Both models predict similar drag coefficients in the linear segment, but the SST-CND model demonstrates better accuracy near stall, aligning more closely with experimental results. In addition, the SST-CND model predicts pitching moment coefficients that correspond more accurately to experimental values than the SST model at angles of attack above 7.05°, as shown in Fig. 19(c).

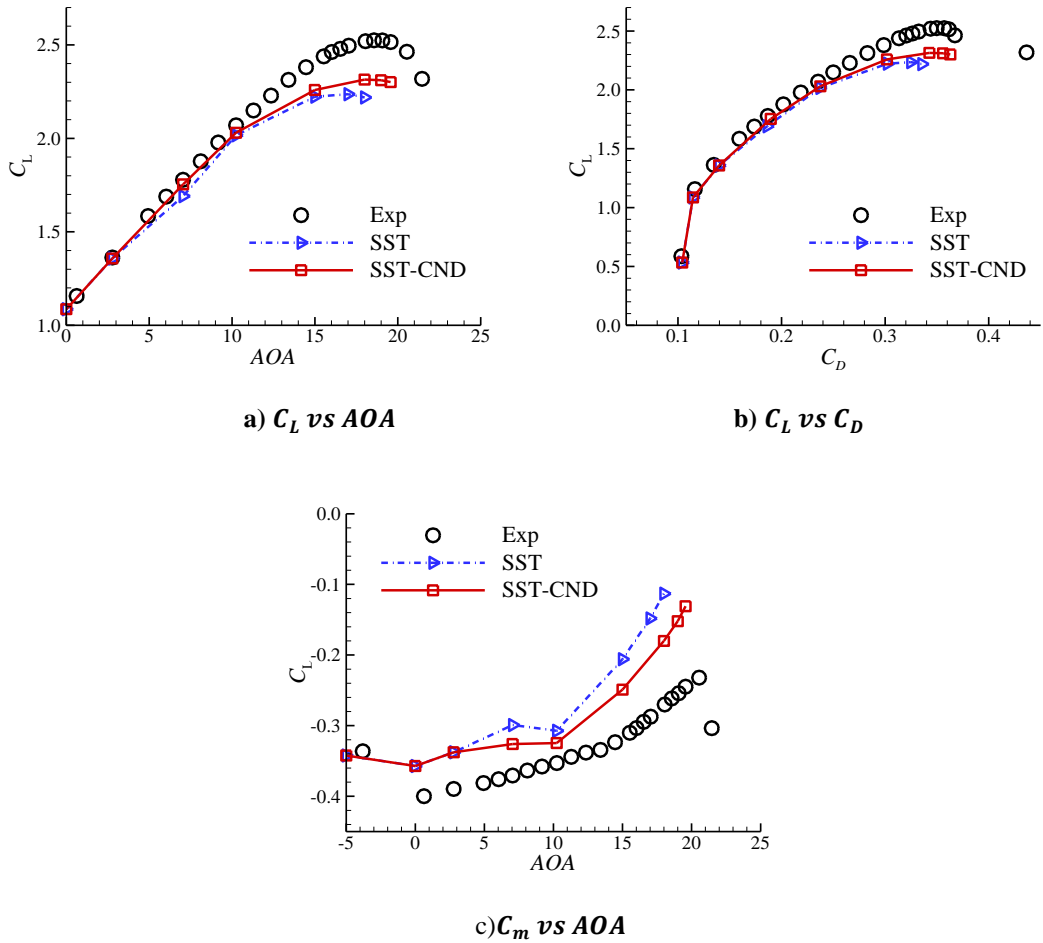

**Fig. 19 Aerodynamic coefficients predicted by the SST and SST-CND models of the CRM-HL configuration**

Fig. 20 illustrates the pressure distributions of the CRM-HL configuration at various span locations at AOA = 7.05°. The SST model underestimates suction peaks, particularly on the flaps along Sections E-E and F-F, and deviates more from experimental data at the main wing's trailing edge in these sections. The SST-CND model demonstrates significant improvement in the wing outboard, predicting more accurate suction peaks in the pressure distribution of the flap. Fig. 21 indicates that the SST-CND model predicts less outboard flap separation compared to the SST model. This reduced separation aligns more closely with the oil flow photograph (Fig. 21a).

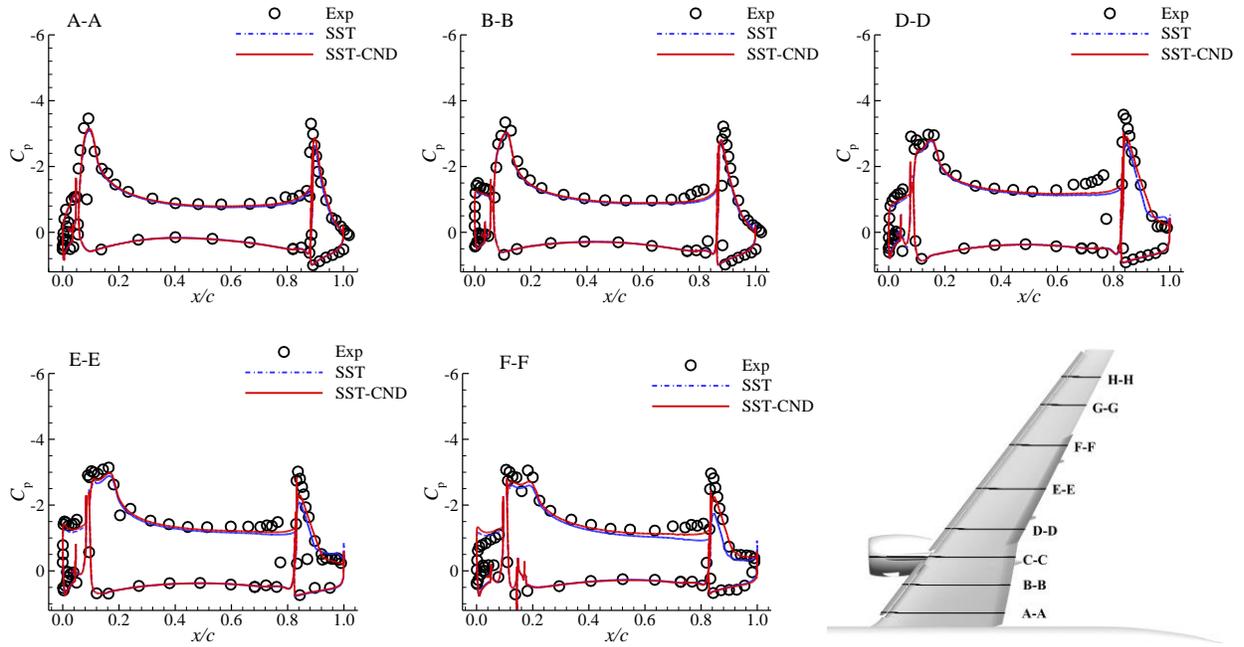

**Fig. 20 Pressure distributions of the CRM-HL configuration at different span locations obtained by different models, AOA = $7.05°$**

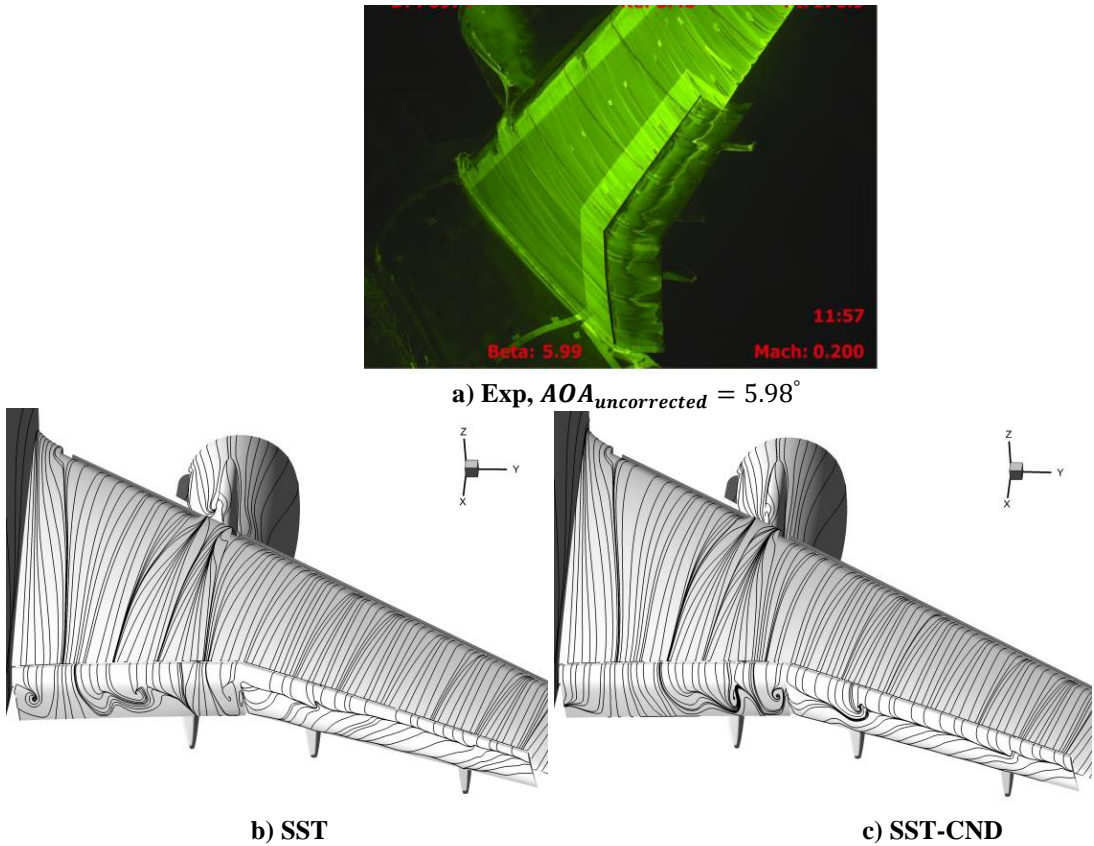

**Fig. 21 Surface streamlines predicted by the SST and SST-CND models and oil flow visualization obtained from the wind-tunnel experiment of CRM-HL, AOA = $7.05°$**

This section explains how the SST-CND model enhances the prediction of flap trailing edge separation. A distinct mixing region forms at the separated shear layer, as depicted in Fig. 22. In this region, the significant increase in $\beta$ enhances momentum diffusion from the mainstream flow to the shear layer, as illustrated in Fig. 22. This leads to a smaller predicted flap trailing edge separation, as illustrated in Fig. 22(b). This phenomenon explains why the SST-CND model predicts a higher flap suction peak and a more accurate lift coefficient.

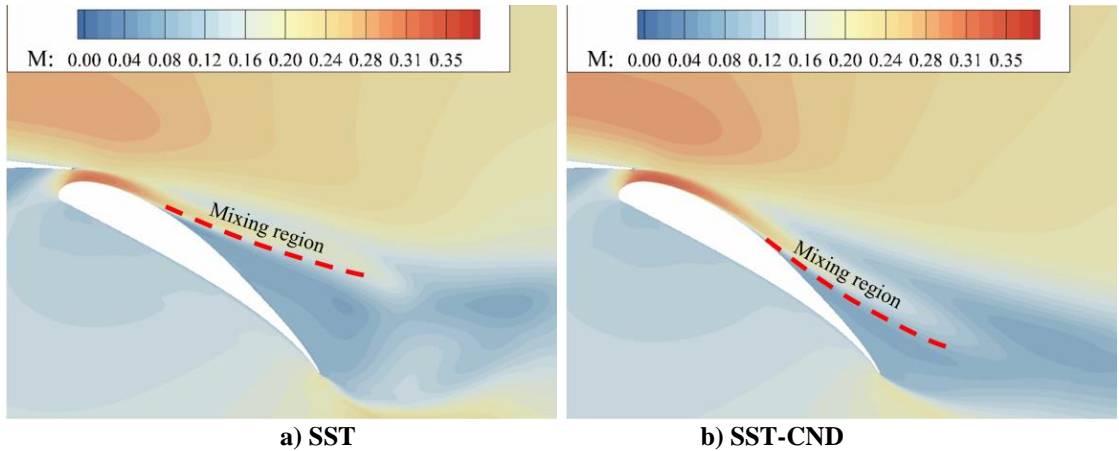

**Fig. 22 Comparisons of the Mach number contours at section E-E obtained using different turbulence models**

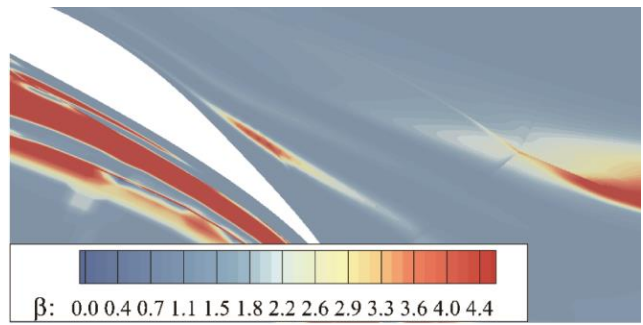

**Fig. 23 $\beta$ distributions at section E-E obtained by the SST-CND model**

Compared to the JSM model, the CRM-HL model exhibits a more complex structure with complex vortices and separated flows near stall conditions. Although the SST-CND model enhances stall prediction for CRM-HL relative to the SST model, the relative error in predicting the maximum lift coefficient remains significant. This discrepancy can stem from two primary factors: first, the CRM-HL configuration operates at a higher Reynolds number (5.49 million), exceeding the training dataset range (approximately 1.0 million); second, the conditional flow field inversion was performed on two-dimensional separated flows, limiting its applicability to complex three-dimensional cases. Future data-driven turbulence models should incorporate training on higher Reynolds numbers and three-dimensional separated flows to improve predictive accuracy.

*3. Results of the $k - \overline{v^2} - \omega - CND$ Model*

A grid convergence study is conducted on the lift, drag, and pitching moment coefficients using the $k - \overline{v^2} - \omega - \text{CND}$ model, as illustrated in Fig. 24. The coarse and medium grids underestimate the maximum lift coefficient, whereas the fine grid shows a slight improvement. The fine and extra-fine grids yield nearly identical results at high angles of attack, closely aligning with experimental data. However, oscillatory convergence becomes more pronounced with the extra-fine grid, as depicted in Fig. 24 (d). The drag coefficient results remain nearly identical across different grids before the stall, while the fine and extra-fine grids provide better agreement with experimental data near the stall. The predicted pitching moment curve more accurately aligns with experimental data as the grid is refined, with the fine grid results effectively capturing the experimental trend. Accordingly, the $k - \overline{v^2} - \omega - \text{CND}$ model demonstrates satisfactory grid convergence in predicting the aerodynamic performance of the CRM-HL configuration. The fine grid is utilized for subsequent simulations to ensure an accurate assessment of stall behavior differences among models.

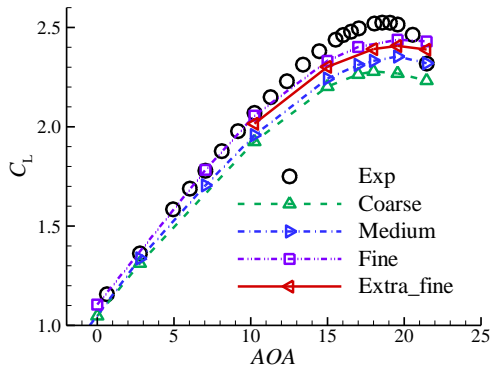

a) $C_L$ vs AOA

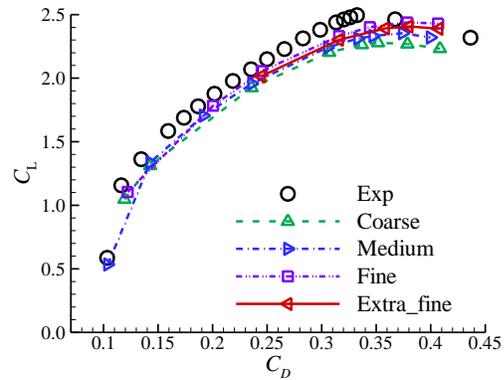

b) $C_L$ vs $C_D$

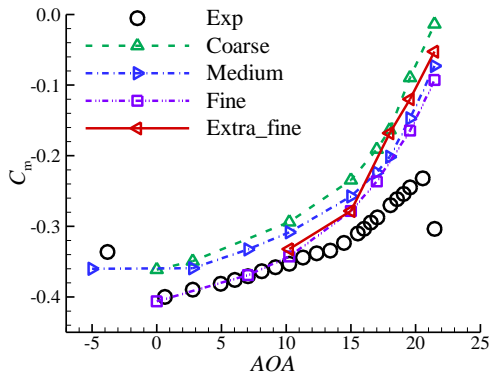
c) $C_m$ vs AOA

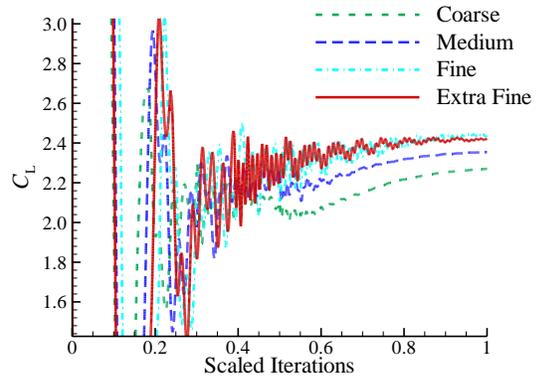
d) Convergence of $C_L$ at $AOA = 19.57°$

**Fig. 24 Mesh convergence of CRM-HL lift coefficient obtained by the $k - \overline{v^2} - \omega - $ CND model**

Fig. 25 presents the aerodynamic performance of the CRM-HL predicted by the $k - \overline{v^2} - \omega$ and $k - \overline{v^2} - \omega$ -CND models using fine grids. The lift coefficients predicted by both models for the linear segment are nearly identical, whereas significant differences are observed under near-stall conditions. The $k - \overline{v^2} - \omega$ model underestimates the maximum lift coefficient by 6.35% and predicts the stall angle to occur 1.57° earlier than the experimental value. In contrast, the $k - \overline{v^2} - \omega - $ CND model significantly enhances prediction accuracy, reducing the relative error of the maximum lift coefficient to 3.17% and delaying the predicted stall angle to 19.57°. It demonstrates good agreement with the experimental data, as detailed in Table 3. In addition, the $k - \overline{v^2} - \omega - $ CND model yields results closer to the experimental data for the polar and pitching moment curves at the near-stall angle of attack, as illustrated in Fig. 25(b) and Fig. 25(c).

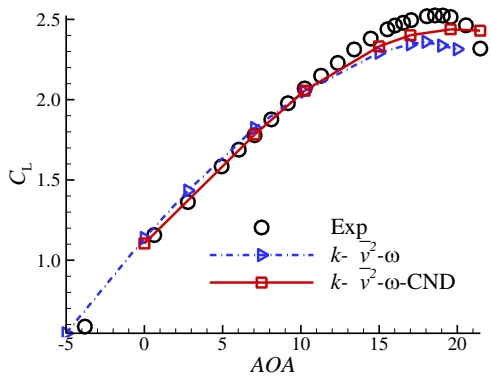
a) $C_L$ vs AOA

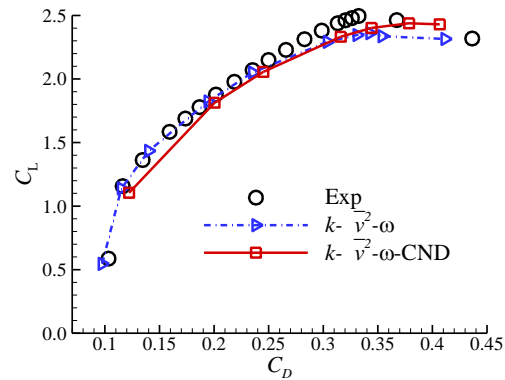
b) $C_L$ vs $C_D$

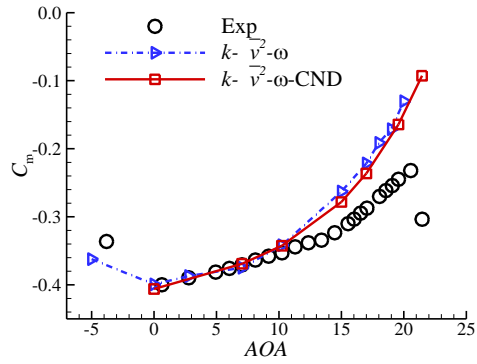

c) $C_m$ vs AOA

Fig. 25 Comparations of the aerodynamic coefficients of the CRM-HL configuration obtained by the $k - \overline{v^2} - \omega$ and $k - \overline{v^2} - \omega -$CND models

Table 3. The results and relative errors of different models

|  | Experiment | $k - \overline{v^2} - \omega$ | $k - \overline{v^2} - \omega -$ CND |
|---|---|---|---|
| $C_{L,max}$/relative error | 2.52/-- | 2.36/6.35% | 2.44/3.17% |
| Stall AOA/deviation | 19.57° /-- | 18° /1° | 19.57° /0° |

Surface flow visualization images are provided in Fig. 26. Fig. 26(a) is derived from oil flow measurements in the wind tunnel test at $AOA_{corrected} = 19.57°$. The presence of the inboard slat cutout induces a distinct wake formation at the root region of the main wing. The wake exhibits an outward inclination toward the outer wing at a specific angle relative to the wing root line, as depicted in Fig. 26(a). Numerical results indicate that the $k - \overline{v^2} - \omega$ turbulence model predicts a wake orientation nearly parallel to the wing root line, as depicted in Fig. 26(c). Comparative analysis indicates that the wake trajectory predicted by the $k - \overline{v^2} - \omega -$ CND model shows significantly improved alignment with the experimentally observed trend, as shown in Fig. 26(d). Therefore, the suction peak of the main wing on the inboard wing predicted by the $k - \overline{v^2} - \omega -$ CND model aligns more closely with the experimental data, as depicted in the A-A and B-B sections of Fig. 27. The oil flow pattern on the outer wing is shown in Fig. 26(b). Vortices from the slat brackets create four distinct separation regions on the outer wing section. The original $k - \overline{v^2} - \omega$ model predicts a relatively large separation on the wing outboard. In contrast, the $k - \overline{v^2} -$ $\omega -$ CND model predicts a smaller separation region, with the pressure distribution at the H-H section of the wing outboard more closely matching the experimental data, as illustrated in Fig. 27.

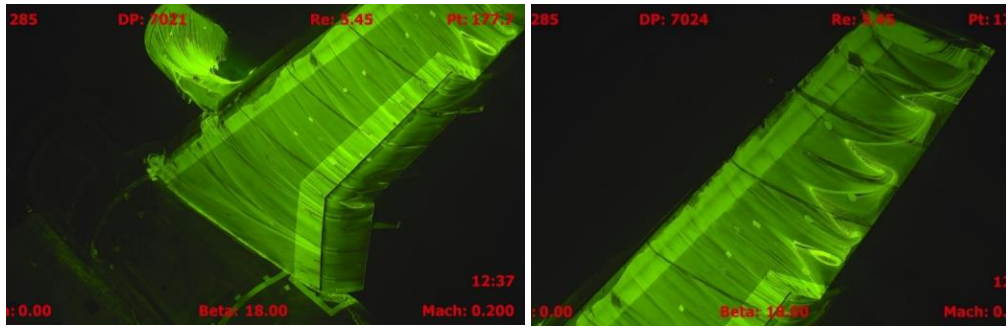

(a) Exp, inboard, $AOA_{corrected} = 19.57°$ (b) Exp, inboard, $AOA_{corrected} = 19.57°$

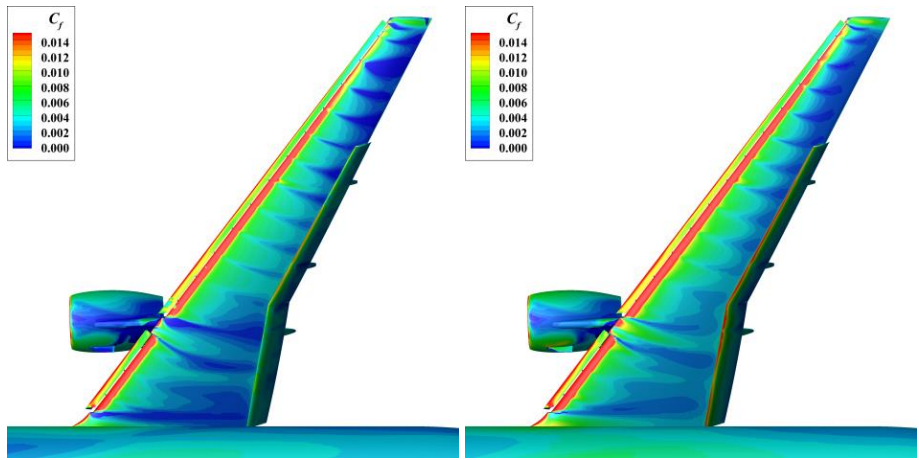

(c) $k - \overline{v^2} - \omega$, AOA $= 19.57°$ (d) $k - \overline{v^2} - \omega -$ CND, AOA $= 19.57°$

**Fig. 26 Surface friction coefficient predicted by the $k - \overline{v^2} - \omega$ and $k - \overline{v^2} - \omega -$ CND models and oil flow visualization obtained from the wind-tunnel experiment of CRM-HL**

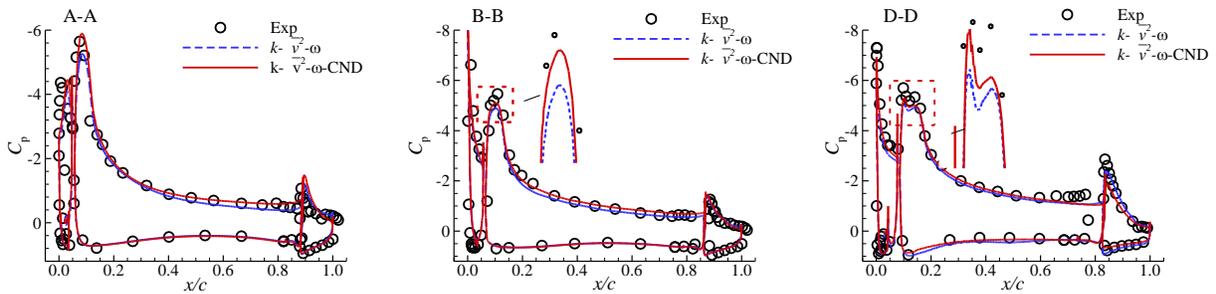

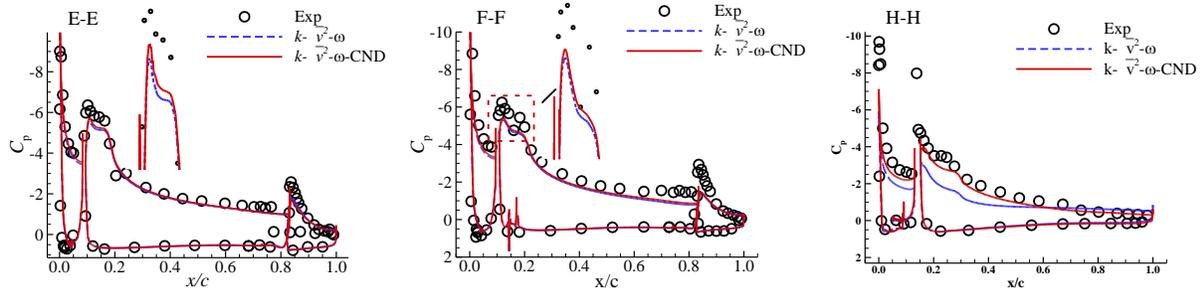

**Fig. 27 Pressure distributions of the CRM-HL configuration at different span locations obtained by different models, AOA = 19.57°**

Fig. 28 presents a comparison of the total pressure coefficient ($C_{pt} = 2(p_t - p_{t\infty})/(\rho V_\infty^2)$) contours for the CRM-HL configuration, as predicted by the $k - \overline{v^2} - \omega$ and $k - \overline{v^2} - \omega - $ CND models. Negative total pressure values indicate a deficiency primarily induced by the vortex system. Vortices originating from the cutout region near the pylon/nacelle for the inboard and outboard slats are referred to as the inboard and outboard vortices, respectively. These vortices merge to form the nacelle/pylon vortex system, as illustrated in Fig. 28(b) and Fig. 28(d). The slat brackets generate distinct separation bubbles on the upper surface of the main wing at high angles of attack, as depicted in Fig. 28 (a) and Fig. 28 (c). The $k - \overline{v^2} - \omega - $ CND model predicts smaller separation bubbles for both the inboard and outboard regions of the wing. This reduction primarily results from a significant increase in $\beta$ within the mixing region between the low-speed vortex core and the mainstream flow. Compared to the baseline $k - \overline{v^2} - \omega$ model, the increase in $\beta$ enhances momentum diffusion between the mainstream and the low-speed vortex core, weakening vortex intensity and accelerating diffusion. Therefore, the $k - \overline{v^2} - \omega$ -CND model predicts smaller separation bubbles on both the inner and outer wing surfaces, leading to improved agreement between predicted and experimental force and moment data. However, the $k - \overline{v^2} - \omega - $ CND model also diminishes the intensity of both the wingtip vortex and the chine vortex. This reduction occurs because the spatially varying correction factor $\beta$, derived from FI-CND, increases in areas with stronger vortices, as shown in Fig. 29. This phenomenon arises due to FI-CND being developed based on two-dimensional separated flow cases. Although this approach ensures accurate prediction of the attached boundary layer, it remains inadequate for fully capturing the characteristics of three-dimensional vortex flows.

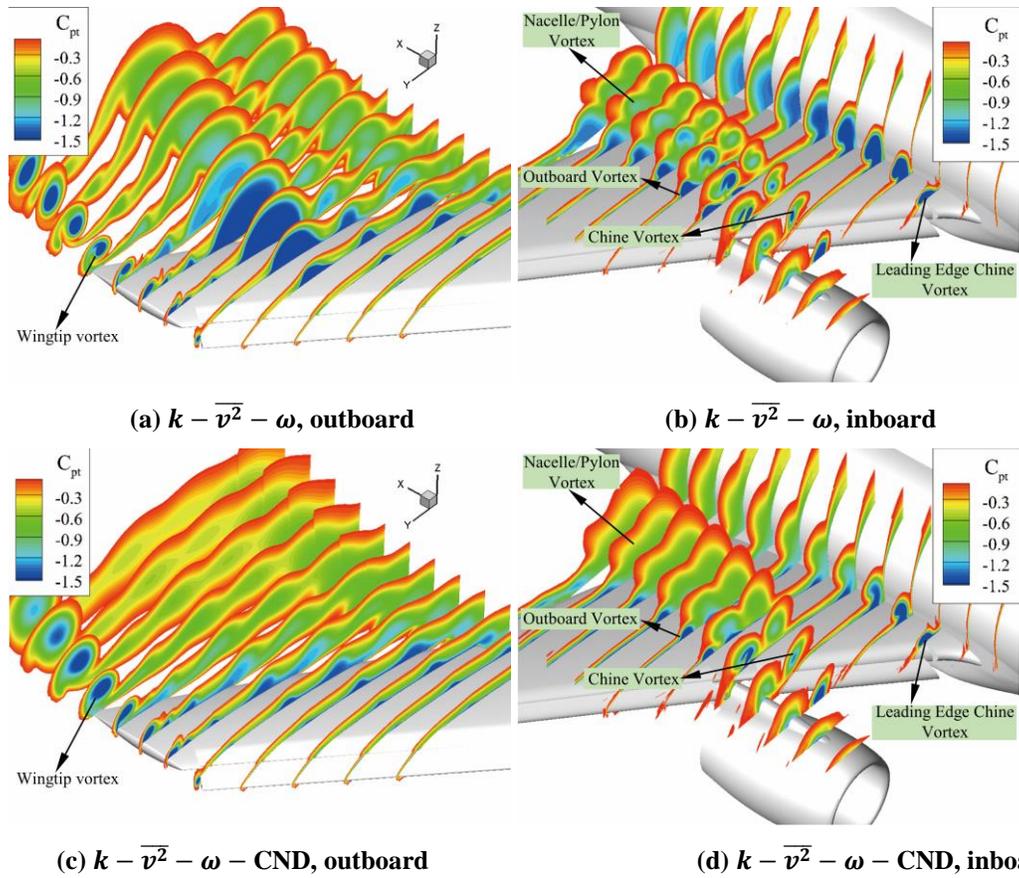

(a) $k - \overline{v^2} - \omega$, outboard  (b) $k - \overline{v^2} - \omega$, inboard

(c) $k - \overline{v^2} - \omega - \text{CND}$, outboard  (d) $k - \overline{v^2} - \omega - \text{CND}$, inboard

**Fig. 28 Comparations of the total pressure coefficient contours of the CRM-HL configuration with and without the nacelle chine, AOA = 19.57°**

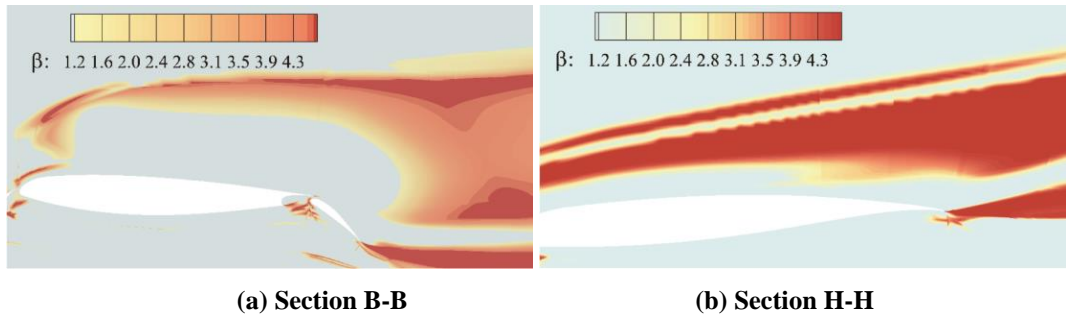

(a) Section B-B  (b) Section H-H

**Fig. 29 $\beta$ distributions obtained by the $k - \overline{v^2} - \omega - \text{CND}$ model, AOA = 19.57°**

## IV. Conclusions

This study examines the engineering applications of data-driven turbulence models. The generalizability of the SST-CND model, derived from conditioned field inversion (FI-CND), is validated for predicting high-lift aerodynamic performance. The correction factor $\beta$ from FI-CND is then applied to the $k - \overline{v^2} - \omega$ model to create the $k - \overline{v^2} - \omega -$ CND model. Numerical simulations are conducted for the 30P30N three-element airfoil, the JAXA Standard Model (JSM), and the high-lift NASA Common Research Model (CRM-HL) are numerically simulated. The work can be summarized as follows.

(1) Numerical simulations using the SST-CND model indicate that it more accurately predicts the stall characteristics of the 30P30N, JSM, and CRM-HL configurations compared to the baseline SST model. The SST-CND model significantly enhances the prediction of flow separation at the flap trailing edge in the linear region. The predicted lift coefficient curve for CRM-HL does not exhibit the pronounced nonlinear rise observed with the SST model. This highlights the strong generalization capability of the SST-CND model in aerodynamic predictions for high-lift configurations.

(2) The baseline $k - \overline{v^2} - \omega$ model demonstrates superior performance over the original SST model in predicting the stall characteristics of CRM-HL. In addition, the $k - \overline{v^2} - \omega -$ CND model improves accuracy in predicting stall behavior. The flow pattern on the wing outboard, as predicted by the $k - \overline{v^2} - \omega -$ CND model, aligns with the experimental data. The relative error in predicting the maximum lift coefficient using the $k - \overline{v^2} - \omega -$ CND model is approximately 3.17% of the experimental data. This result highlights the strong transferability of model corrections derived from conditioned field inversion across different turbulence models.

(3) The model derived from FI-CND exhibits good generalization in predicting stall characteristics of high-lift configurations. However, it also induces the adverse effect of weakening the wingtip vortices and chine vortices. Future data-driven turbulence modeling efforts should incorporate additional three-dimensional physical mechanisms and be conducted under higher Reynolds number conditions.

## Acknowledgments

This work was supported by the National Natural Science Foundation of China (grant nos. 12372288, 12388101, U23A2069, and 92152301).